\documentstyle[twocolumn]{article}
\makeatletter
\def\@normalsize{\@setsize\normalsize{12pt}\xpt\@xpt
\abovedisplayskip 10pt plus2pt minus5pt\belowdisplayskip \abovedisplayskip
\abovedisplayshortskip \z@ plus3pt\belowdisplayshortskip 6pt plus3pt
minus3pt\let\@listi\@listI}

\def\subsize{\@setsize\subsize{12pt}\xipt\@xipt}

\def\section{\@startsection {section}{1}{\z@}{24pt plus 2pt minus 2pt}
{12pt plus 2pt minus 2pt}{\large\bf}}

\def\subsection{\@startsection {subsection}{2}{\z@}{12pt plus 2pt minus 2pt}
{12pt plus 2pt minus 2pt}{\subsize\bf}}
\makeatother

\begin{document}

%don't want date printed
\date{}

%make title bold and 14 pt font (Latex default is non-bold, 16 pt)
\title{\Large\bf CAPANIC: A Parallel Tree N-body code for inhomogeneous
clusters of processors.}

%for single author (just remove % characters)
%\author{I. M. Author \\
%  My Department \\
%  My Institute \\
%  My City, ST, zip}

%for two authors (this is what is printed)
\author{\begin{tabular}[t]{c@{\extracolsep{8em}}c}
  V. Antonuccio-Delogu	& U. Becciani\\
 \\
  Catania Astrophysical Observatory & Catania Astrophysical Observatory \\
  Citt\'{a} Universitaria & Citt\'{a} Universitaria\\
  Catania, ITALY~~I-95125	 & Catania, ITALY~~I-95125
\end{tabular}}

\maketitle

%I don't know why I have to reset thispagesyle, but otherwise get page numbers
\thispagestyle{empty}

\subsection*{\centering Abstract}
%IEEE allows italicized abstract
{\em
We have implemented a parallel version of the Barnes-Hut 3-D
N-body tree algorithm under PVM 3.2.5, adopting an SPMD paradigm.  We
parallelize the problem by decomposing the physical domain by means of the
{\bf Orthogonal Recursive Bisection} oct-tree scheme suggested by Salmon
(1991), but we modify the original hypercube communication pattern into an
incomplete hypercube, which is more suitable for a generic inhomogenous
cluster architecture.\\
We address dynamical load balancing by assigning different "weights" to
the spawned tasks according to the dynamically changing workloads of each
task. The weights are determined by monitoring the local platforms where
the tasks are running and estimating the performance of each task. The
monitoring scheme is flexible and allows us to address at the same time
cluster and intrinsic sources of load imbalance. We then show
measurements of the performance of our code on a test case of astrophysical
interest in order to test the performance of our load-balancing scheme.
%end italics mode
}

\section{Introduction}

Numerical simulations of gravitationally interacting particles have become one
of the
most powerful tool of contemporary cosmology.  Objects ranging in size from
stellar globular clusters up to clusters of galaxies, and including elliptical
and spiral
galaxies, can be regarded
as made of a large collection of point-like particles interacting through the
Newton's
law of gravitation. The {\bf Gravitational N-body Problem} aims at finding a
description of the dynamics of such systems of $N$ gravitationally interacting
particles by solving  directly their equations of motion:

\begin{equation}
m_{i}\frac{d^{2}{\bf r}_{i}}{dt^{2}}=-\sum_{j\neq i, j=1}^{N}\frac{Gm_{i}m_{j}}
{\mid{\bf r}_{i}-{\bf r}_{j}\mid^{3}}\left({\bf
r}_{i}-{\bf r}_{j}\right)   \label{eq1}
\end{equation}

In the above equations $G$ denotes Newton's constant, $m_{i}$ is the mass of
$i$-th
particle and the index $i$ runs form 1 to N.\\
These equations describe the dynamics of a system of point-like particles
interacting
only through their mutual gravitational force. Although stars within a globular
cluster
or a galaxy within a galaxy cluster are {\em extended} objects, they can be
considered
point-like as far as the density and the typical velocity are enough small as
to make
 the probability of close interactions (which could destroy individual objects
and then
create new objects within the system) very small. Back-of-the-envelope
calculations
show that this
condition is fulfilled at least for typical galaxies and clusters of galaxies,
so that
eq.(~\ref{eq1}) provides a reasonable description of the system.The
gravitational
N-body problem provides then a good model of  these systems, although it does
not
take
into account the presence of gas, which is observed within galaxies and
clusters of
galaxies. However, one can show that the gas can significantly affect the
dynamics of
galaxies only after they have formed, and has little influence in determining
the
global observed properties of clusters of galaxies. Under many respects,
eq.(~\ref{eq1})
provides an accurate description of the objects which trace the Large Scale
Structure of the Universe. This is even more believable when one considers
that,
according to modern cosmology, $90$ \% of the total gravitating mass of the
Universe
should be made of particles which do not emit an appreciable amount of
radiation at
any wavelength and interact with the visible matter which makes up stars and
galaxies only through their gravitational action. This {\bf Dark Matter}, if
it exists, dominates the mass content of the Universe, and its dynamics is
exactly
described by eq.(~\ref{eq1}).\\
System of equations similar to  eq.(~\ref{eq1}) are encountered in other fields
of
physics, e.g. in plasma dynamics and in molecular dynamics. However, in the
latter
case the interaction force decays usually faster than $r^{-2}$ with distance,
so that
one commits a small error in considering the interaction as a local one. In
plasma
dynamics the product $m_{i}m_{j}$ in eq.(~\ref{eq1}) is replaced by the product
of the
charges $q_{i}q_{j}$, and in stationary neutral plasma the repulsive and
attractive
interaction balance in such a way that on scales larger than a ``Debye radius''
the
plasma can be considered as neutral. This fact simplifies enormously the
calculations: in fact one needs to consider only the interactions of a particle
with
those particles lying approximaztely within a Debye radius in order to give a
detailed
account of the collective dynamics of a plasma. This however does not hold for
the
gravitational force: being always attractive, and decaying not enough fast with
distance, the gravitational interaction does not allow any ``screening''. The
analogous
of the Debye radius for gravitating system does not exist: the contribution
to the gravitational force from a large collection of far bodies is on average
quantitatively as important as that coming from random enconuters with nearby
bodies. These facts all influence the choice of the correct integration methods
of
the above equations.\\
The use of N-body computer codes to find approximate numerical solutions of the
eqs. (~\ref{eq1}) has become a primary tool of the  theoretical research in
cosmology.
The simplest and oldest algorithm
was based on a direct solution of the equations for each body (\cite{aar:foo}).
One
sees immediately that the computer work will increase as $O(N^{2})$ in this
method, a cost which makes prohibitive simulations of more than about $10.000$
particles even on the largest present-day parallel machines. This difficulty
prompted
for the search of alternative algorithms which could circumvent this problem.
Eq.(~\ref{eq1}) can be rewritten in terms of multipole expansion as:
\begin{equation}
m_{i}\frac{d^{2}{\bf r}}{dt^{2}}=-\frac{Gm_{i}M}{\mid{\bf r}_{cm}\mid^{3}}
{\bf r}_{cm} +
\frac{{\bf Q}^{ij}\left({\bf r}-{\bf r}_{cm}\right)_{i}
\left({\bf r}-{\bf r}_{cm}\right)_{j}}{\mid{\bf r}-{\bf r}_{cm}\mid^{5}} + ...
\label{eq2}
\end{equation}
In this formula $M$ is the total mass of the system, and a subscript $cm$ means
that the given quantity is computed at the {\em center of mass} of the system.
The second term on the right-hand side is the {\em quadrupole term},
 a set of 6 quantities obtained from integrals over the system, and we have
omitted to report the higher order terms. The summation over all the particles
in
eq.(~\ref{eq1}) has now been replaced by a summation over the multipole
expansion.
Although these latter are infinite in number, their magnitude is a fast
decreasing
function of distance from the center of mass, so we commit a little error by
omitting
the higher order terms in this approximation.\\
The use of eq.(~\ref{eq2}) to compute the force is at the heart of the
Particle-Mesh
$(PM)$ and (Particle-Particle)-(Particle-Mesh) $(P^{3}M)$ numerical approaches
(see \cite{h:e} for a comprehensive treatment). In the PM mehod the
gravitational
potential and
force are calculated from a Fast Fourier Transform of the density distribution,
and
the force is computed at the corners of a grid superimposed on the system. The
spatial resolution is rigidly fixed by the size of the mesh, and cannot be
changed
during the simulation. The approximation then becomes a poor one as the system
evolves toward an inhomogeneous situation, as it often happens in cosmological
simulations in which clusters of galaxies form out of an almost uniform initial
state.
In the $P^{3}M$  the interaction with the nearest neighbouring particles is
computed
exactly,  but otherwyse the computation scheme is that of a $PM$. In both these
schemes the computational cost grows at most as $O(N\log N)$, but the grids are
fixed
and the final spatial resolution depends on their size, i.e. on the number of
particles
adopted. Another multipole expansion scheme was devised
by Greengard and Rokhlin \cite{g:r}.\\
An adaptive algorithm class of multipole expansion methods is based on the
{\em tree decomposition} to represent the structure of the gravitational
interactions. Our code is based on this algorithm, and in the next section we
will
look at it in more detail.

\section{Tree Methods.}
In the Barnes-Hut tree  method \cite{b:h} the space domain containing the
system is
divided into a set of cubic cells by means of an oct-tree decomposition:
starting from
a root cell
containing all the particles each cell is further subdivided into 8 cells,
until the last
cells contain only 1 or 0 bodies. This structure is the {\em tree}. For those
cells of the
tree
containing more than one body one stores the position, size, total mass and
quadrupole moment in corresponding arrays. Cells containing only one
body, on the other hand, store only the position of the body. To compute the
force on a given particle, one inspects the tree, i.e. compare the particle's
position with the distance and position of each cell of the tree. When the
distance
$r$ of the particle from the edge of the given cell is such that the {\em Cell
Opening Criterion (COC)} is verified, i.e. when: $r/d<\theta$, where d is
the size of the edge of the cell and $\theta$ is a fixed parameter, the cell is
``opened''.
i.e. one inspects the cells which
are ``daughters'' of the current cell. Cells containing only one body and cells
which do
not fulfill the {\em COC} are considered for interaction: their monopole and
quadrupole moments (if they have one) are added up to the total gravitational
force
felt by the body. In this way, the interaction with nearest single particles is
computed exactly, while groups of particles which are far enough treated as
extended objects characterized by a monopole and a quadrupole interaction.\\
The Barnes-Hut scheme has been implemented in various numerical N-body codes.
It is adaptive (the tree is reconstructed after each time step) and through the
parameter $\theta$ allows a control on the accuracy of the force calculation,
ranging
from the case $\theta=0$ (corresponding to direct interactions) to larger
values.
The parallel PVM N-body code which we have developed is based on the FORTRAN
version of the vectorial code written by dr. L. Hernquist \cite{h}, who kindly
provided
us with a
copy of his latest version. However, as we will show later, the communication
structure of our CAPANIC code bears little resemblance with that of the
original
Hernquist's code.\\
A parallel implementation of the Barnes-Hut algorithm was made by Salmon
\cite{s},
\cite{w:s}, \cite{s:w}, \cite{s:w:w}, \cite{f:q:g:s:w}. It was devised to run
on massively
parallel systems like the CM-5 and the Touchstone Delta, and the
parallelization was
done exploiting features of the FORTRAN compilers on these machines. Another
implementation on a dedicated, transputer -based  machine, (GRAPE 1-A) was made
by Makino and coll. \cite{m:1}, \cite{f:i:m:e:s}. Although the results are very
encouraging, these codes are difficul to export on platforms different from
those for
which they were originally devised. Moreover, they are devised for {\em
homogeneous} clusters of processors, where all the processors have the same
characteristics, although some of them account for dynamical load balancing
among different processors.\\
The strategy behind our parallel N-body code is complementary to that of the
above
quoted papers. We desired to produce a public software which could be easily
implemented on different platforms, and particularly on already existing
clusters of
workstations. It was then necessary to perform the parallelization by adopting
products which are (or start to be viewed as) standards, and
which can be readily obtained by everybody. This motivated our choice of making
use of the  Parallel Virtual Machine ({\bf PVM }) software, which is now
emerging as a
standard and is particularly suited to implement parallel algorithms on
clusters of
heterogeneous workstations.\\
When one writes a parallel application one has to make a choice between two
possible schemes: a ``master/slave'' and a ``Single Program Multiple Data''
(SPMD) one.
In the latter a single program is ``cloned'' within the tasks spawned by the
initial
application, and the cloned parts control the communication of data among the
different processes. We adopted this latter scheme because all the processors
are
treated on an equal foot, while in the master/slave scheme the processor which
hosts
the master process does a different job, and generally it spends a long time
waiting
for the data to come back from the slave processes. On the other hand, an SPMD
scheme, although is generally more difficul to implement, is more suited for
dynamical load balancing, one of the crucial issues addressed in our work.\\
The result of this effort is {\bf CAPANIC} ({\bf
CA}tania {\bf PA}rallel {\bf N}-body Code for {\bf I}nhomogeneous {\bf
C}lusters of Workstations),
which described in the following sections.

\section{PVM and parallelism.}

\subsection{Overview.}
CAPANIC has been devised to work under PVM (Parallel Virtual Machine), a
package freely distributed by the Oak Ridge National Laboratory, Tennessee. We
run
it on a cluster comprising a Convex C210 and various Sun workstations at our
Institute.

\subsection{PVM short description.}

PVM is based on messages passing between heterogeneous hosts on the network.
It creates on each host belonging to the virtual machine a daemon which is able
to
send/receive messages and data packets to/from other daemons loaded on other
hosts. In this way a collection of different processors can be seen by a
specified job
as a large, single virtual machine.\\
PVM automatically start up tasks on the hosts and provide a suitable set of
subroutines for message
passing in order to allow the tasks to communicate and synchronize with each
other. Applications
written in Fortran77 and C can be parallelized using the PVM message passing
contructs, so that
several communicating tasks can cooperate to solve the problem.\\
Each process enrolling in PVM is assigned an integer task identifier (tid) that
unambigously identifies it. The tids are unique across the virtual machine and
are
provided by the PVM daemons.\\
PVM supplies routines for packing and sending messages and data among tasks on
the virtual machine. Each message of the transmitting task is packed on a
send-buffer on the host and sent
to the receive-buffer of the receiving tasks. The communication model provides
asynchronous
blocking send functions that returns when the send buffer is free for reuse
(reception is completed on
the receive buffer of the receiving tasks); asynchronous blocking receive
function that returns when
the data are received in the buffer; and asynchronous non-blocking receive
function that returns with
either the data or a flag that data has not arrived.\\
PVM supports point-to-point communication and multicast to a set of taks and
broadcast to a
user defined group of tasks. Message buffer are allocated dinamically, so that
the maximum size
messages that can be sent or received depends only on the available memory of
the hosts.

\section{CAPANIC,  inhomogeneous clusters and load balancing}

The CAPANIC software has been devised to run on an inhomogeneous cluster of
hosts forming the
virtual machine. At our site the hosts are used as general purpose computers
and
their local load can be strongly variable with time.\\
At the beginning the code divides the physical domain occupied by the system
into an incomplete hypercube, and assigns the bodies contained within M regions
of
the hypercube to an equal amount of tasks which are spawned on the hosts of the
virtual machine. In order to avoid "load imbalance" the domain decomposition
is performed taking into account the fact that the total workload on each task
depends on two different types of parameters. Parameters of the first type,
$A_{i}$
(i=1..M), depend on the characteristics of the hosts
where the tasks will be spawned; those of the second type, $B_{j}$ (j=1..N),
depend
on the number of
interactions that are necessary to calculate the force on the body. The $A_{i}$
parameters are evaluated
from the statistical average load and the characteristics of the host , and
increase
with the average performances associated with the host. The $B_{j}$ parameter
are
evaluated using information from previous runs, and account for the intrinsic
work
done by the spawned tasks to advance in time the positions and velocities of
the particles which have been assigned to it. This work depends strongly on
the structure and depth of the tree: ultimately on the geometry and mass
distribution
of the particles within the system. At the beginning we put $B_{j}=1$ for all
the
particles.

\subsection{The domain decomposition}

Following Salmon \cite{s} we use an orthogonal recursive bisection (ORB) scheme
to partition the entire domain in M subdomains and
to assign the particles of each partition to a task. Each cutting plane
(bisector) of the
partition splits the domain into two
subdomains  to which a set of processors is assigned. The domain decomposition
proceeds until only one processor is assigned to each subdomain.
The position of each bisector is determined in such a way as to have the same
workload in each of the subdomains.\\
Let us introduce the function $W(x)$ defined as the
ratio of the works associated
with the subdomain and the work associated with the parent domain:
\begin{equation}
W(x)=\frac{Work({\rm subdomain})}{Work({\rm parent~domain})}
\end{equation}
where the Work function is evaluated as the following ratio
\begin{equation}
Work({\rm region})=\frac{\sum_{1}^{N_{bodies}}B_{j}}{\sum_{1}^{N_{proc}}A_{i}}
\end{equation}
where $j=1,..., N_{bodies}$ runs over the bodies of the region, and
$i=1,..., N_{proc}$
 is the total number of processors assigned to
the region. The position of the bisector $x_{split}$ is choosen so that
$W(x_{split}) =
0.5$. After the domain decomposition, the tasks are spawned on the virtual
machine,
and the properties of the bodies
of the subdomain assigned to each task are delivered, so that each task contain
only
the right amount of information needed to compute the forces on the bodies
assigned
to it.

\section{Results.}

The virtual machine we have used to develop and test our application is formed
by the
following hosts: one Convex C210  machine having 64 MB Ram, one Sun Sparc 10
(first generation) having 32 MB Ram, one Sun Sparc 10 (last generation)
having 32 MB Ram; three Sun Sparc 2 having 16 MB Ram, one Sun Sparclassic
having
16 MB Ram.
	Runs were realized in order to compare the efficiency of the CAPANIC code
on the virtual machine, in comparison with the vectorial code of Hernquist.
	Several tests were realized for a system of 8000 bodies
in a configuration evolving slowly
for about 20 time-step.  It is useful to distiguish 4 phases at each time-step:
1- local
tree formation (computational phase); 2 - send/receive trees (communication
phase); 3 - locally
essential tree formation, force evaluation on each local body, update bodies
proprierties
(computational phase); 4 - body migration and synchronization (communication
phase).\\
We reports only the results of the most significant tests.

\subsection{First Test}
Here we run the serial Hernquist's code on CONVEX machine with, on average,
40\% CPU, so
that the time-step duration was about 55 sec. Using a full dedicated Sun Sparc
2 the
time-step duration was about 172  sec.

\subsection{Second Test}
Using CAPANIC we performed three runs using  2, 4 and 8  tasks spawned on the
CONVEX, with 4000, 2000 and 1000 bodies for each task, respectively.
The following figure show the results.

\vspace*{10cm}
As reported in the figure, we can distinguish four phases.
In the A phase, using only the local bodies, the tasks form a local tree.
The  computational time spent
in this phase depends on two factors: the  number of
bodies assigned to each task, which decreases as the
number of spawned tasks increase, and the total CPU time, that increase as the
number of spawned task increase, although the CPU time for each task decrease
from 35\%  (2 spawned task) to 11\% (8 tasks).\\
In  the B phase, the tasks send and receive the information from each other
task of
the application.
For each bisector of the splitted domain the cells of the local tree in each
task are
checked: those which do not satisfy a {\em Domain Opening criterion [DOC]}
(\cite{s}) are enqueued for sending, while those which satisfy the {\em DOC}
are
opened and their daughters are inspected. Then the task sends the cell
properties to
the processor set on the other side of the
bisectors. After this step,it receives cell properties from all the tasks of
the
application. During
this phase the computational time is negligible and we can consider this time
to be
dominated by the time of the communication
phase.\\ In the C phase, using the received cells, the task builds the locally
essential tree
(\cite{s}). For each local body, traversing the locally essential tree, the
force acting on
it  is evaluated and body properties are updated. The computational time spent
in this
phase depends on the same factors of the A phase.\\
In the D phase, local bodies that are out of the spatial region assigned to the
task, are
deleted from
the list of local bodies and delivered to the tasks that have the spatial
region
including the new bodies
position (body migration phase). At the same time a task receives bodies from
the
other task.  During this
phase the computational time is negligible and we can consider this time to be
dominated by the communication phase.\\
We can note that the communication and synchronization phases (B and D phases)
increase from
2.3\% of the total time-step (2 tasks) to 38.6\% (8 tasks). This is essentially
due to the
load of the PVM daemon that serves all the pvm calls, whereas the time for the
computational phase decrease from
97.7\% to 61.4\% of the total time step. The time step duration decrease from
35.03 sec
 to about 40 sec as the number of spawned tasks on the same host increases.

\subsection{Third Test}
We have spawned 2 tasks: 1 task on Convex C210 host (using 45\% CPU time) with
6300 bodies and
1 task on Sun Sparc 2 host (using 90\% CPU time) with 1700 bodies. We have
verified
that with this choice the workload is approximately the same in both tasks, and
we
obtain the results shown in the following figure:

\vspace*{10cm}

We can note that this test produces results similar to the previous one using 2
task
spawned on the Convex. The time spent in the B and D phases $T(comm)$ depends
on three
factors: $T(l)$, the latency time due to the PVM software, $T(n)$ the network
transfer
time and $T(s)$, the waiting time for synchronization. We have:
\begin{equation}
T(comm)=T(l)+T(n)+T(s)
\end{equation}
During the run the total network capability was used for the communication
phase.
For the task spawned on the Convex  the following results were obtained:
\begin{center}
T(l)=T(n);    T(l)=0.17 sec; T(s) was negligible
\end{center}
For the task spawned on Sun Sparc 10 host the following data were evaluated:
\begin{equation}
T(l)=T(n) ;   T(l)=0.17 sec;    T(s)=8.9*T(n)\approx 1.51 sec
\end{equation}

\subsection{Fourth Test}

We have spawned 7 tasks: 1 task on Convex machine using 50\% CPU time with 2462
bodies; 1 task
for each Sun Station of the virtual machine (full dedicated to run the tasks).
The Sun Sparclassic and
the Sun Sparc2 machines with 667 bodies (667 x 4 = 2668 bodies); the Sun Sparc
10 (first
generation) with 1230 bodies and the Sun Sparc 10 (last generation) 1640
bodies. Using this domain
distribution the load was alomost equally balanced between the tasks, and we
obtain the following results:

\vspace*{10cm}

It is interesting to observe that, in comparison with the second test using 8
task on Convex, the communication
and synchro phase decrease from 15.3 sec. to 6.3 sec
(considering the slowest machines), and the following results for the speed-up
S were
obtained:
\begin{equation}
S_{1}=\frac{T({\rm serial~code~on~Convex})}{T({\rm 7~spawned~tasks~on~the~
virtual~machine})}=1.98
\end{equation}
and
\begin{equation}
S_{2}=\frac{T({\rm serial~code~on~SunSparc~2})}{T({\rm 7~spawned~tasks~on~the~
virtual~machine})}=6.2
\end{equation}

where $T(machine)$ is the time step duration.\\
In the CAPANIC application $T(machine)$ can be considered as the sum of the
computational time $T(comp)$ and the communication and synchronization phase
$T(comm)$. Depending on the host were
the task was running $T(comm)=T(l)+T(n)+T(s)$ we derive the following values:
\begin{center}
Sun Sparcstation 2: T(s) negligible;\\  T(n)= 2*T(l) = 4.2 sec.
\end{center}
\begin{center}
Sun Sparcstation 10: T(s) =5.2 sec;\\  T(n)= 2*T(l) = 4.2 sec.
\end{center}
\begin{center}
Convex C210: T(s)= 7 sec;\\  T(n)= 2*T(l) = 4.2 sec.
\end{center}

\section{Summary and Future Prospects.}
The results presented in the preceding section demonstrate the efficiency of
the PVM
package in parallelizing an highly adaptive, dynamically changing alogorithm
like the
Barnes-Hut tree application. Our tests show that the total speed-up rises to
approximately 6.2 for our inhomogeneous cluster. It is true that we evolved our
system only for 20 time steps, so our system had not enough time to become very
inhomogeneous, but it has been observed that the load does not depend
very much on the degree of inhomogeneity of the system (\cite{w:q:s:z}).
Probably the
total number of particles is a more crucial parameter, and we plan to test
CAPANIC
for initial configurations having $N>8000$.\\
We are now planning to extend the original tree algorithm to include the gas.
As we
said in the Introduction, this component does not affect significantly the
dynamics on
scales larger than those of individua galaxies. However it is the component
which
emits most of the visible light in the Universe, so it is important  to
introduce it within
a code designed for cosmological simulations. An obvious choice would be to
extend our CAPANIC code by ``merging'' it with an SPH code, but recently Motta
and
Wick (\cite{m:w}) have proposed a particle approximation scheme to solve
numerically the
Fokker-Planck equation which is far more accurate and simple to implement than
the SPH. At variance with the SPH the Motta-Wick algorithm is based on an
approximation to an exact solution of the kinetic equations, where the fluid
elements
are treated as gas particles having a dynamics specified by a set of
gravitational
equations and fluid equations. The main problem lies in the fact that the
introduction of a second class of particles brings cell-cell communications
into the
game, a feature which complicates the communications. We will report on this
attempt in another paper.

\end{document}